\documentclass{IOS-Book-Article}

\usepackage{mathptmx}
\usepackage{graphicx}
\usepackage{amsmath}
\usepackage{epstopdf}
\usepackage{subfigure}

\newcommand\blfootnote[1]{%
  \begingroup
  \renewcommand\thefootnote{}\footnote{#1}%
  \addtocounter{footnote}{-1}%
  \endgroup
}

\begin{document}

\begin{frontmatter}

\title{A Survey of Interdependency Models for Critical Infrastructure Networks}
\author[A]{\fnms{Joydeep} \snm{Banerjee}%
\thanks{Correspondence To: Joydeep Banerjee, E-mail: jbanerje@asu.edu}},
\author[A]{\fnms{Arun} \snm{Das}%
\thanks{Correspondence To: Arun Das, E-mail: adas@asu.edu}}
and
\author[A]{\fnms{Arunabha} \snm{Sen}%
\thanks{Correspondence To: Arunabha Sen, E-mail: asen@asu.edu}
}

\runningauthor{J. Banerjee et al.}
\address[A]{Computer Science and Engineering Program\\
School of Computing, Informatics and Decision System Engineering\\
Arizona State University, Tempe, Arizona 85287, USA}

\begin{abstract}
The critical infrastructures of the nation such as the power grid and the communication network are highly interdependent. Also, it has been observed that there exists complex interdependent relationships between individual entities of the power grid and the communication network that further obfuscates the analysis, and mitigation of faults in such multi-layered networks. In recent years, the research community has made significant efforts towards gaining insight and understanding of the interdependency relations in such multi-layered networks, and accordingly, a number of models have been proposed and analyzed towards realizing this goal.

In this chapter we study existing interdependency models proposed in the recent literature and discuss their approach, and inherent features, towards modeling interdependent multi-layer networks. We also provide a brief discussion into the drawbacks of each of these models and propose an alternate model that addresses these drawbacks by capturing the interdependency relationships using a combination of conjunctive and disjunctive relations.
\end{abstract}

\end{frontmatter}

\section{Introduction}
\blfootnote{Acknowledgment: This research was supported in part by the DTRA grant HDTRA1-09-1-0032 and the AFOSR grant FA9550-09-1-0120. The Maricopa county communication network data used in this research was provided by GeoTel communications (www.geo-tel.com).}

\begin{figure}[h]
\label{triangle}
  \centering
    \includegraphics[width=0.75\linewidth]{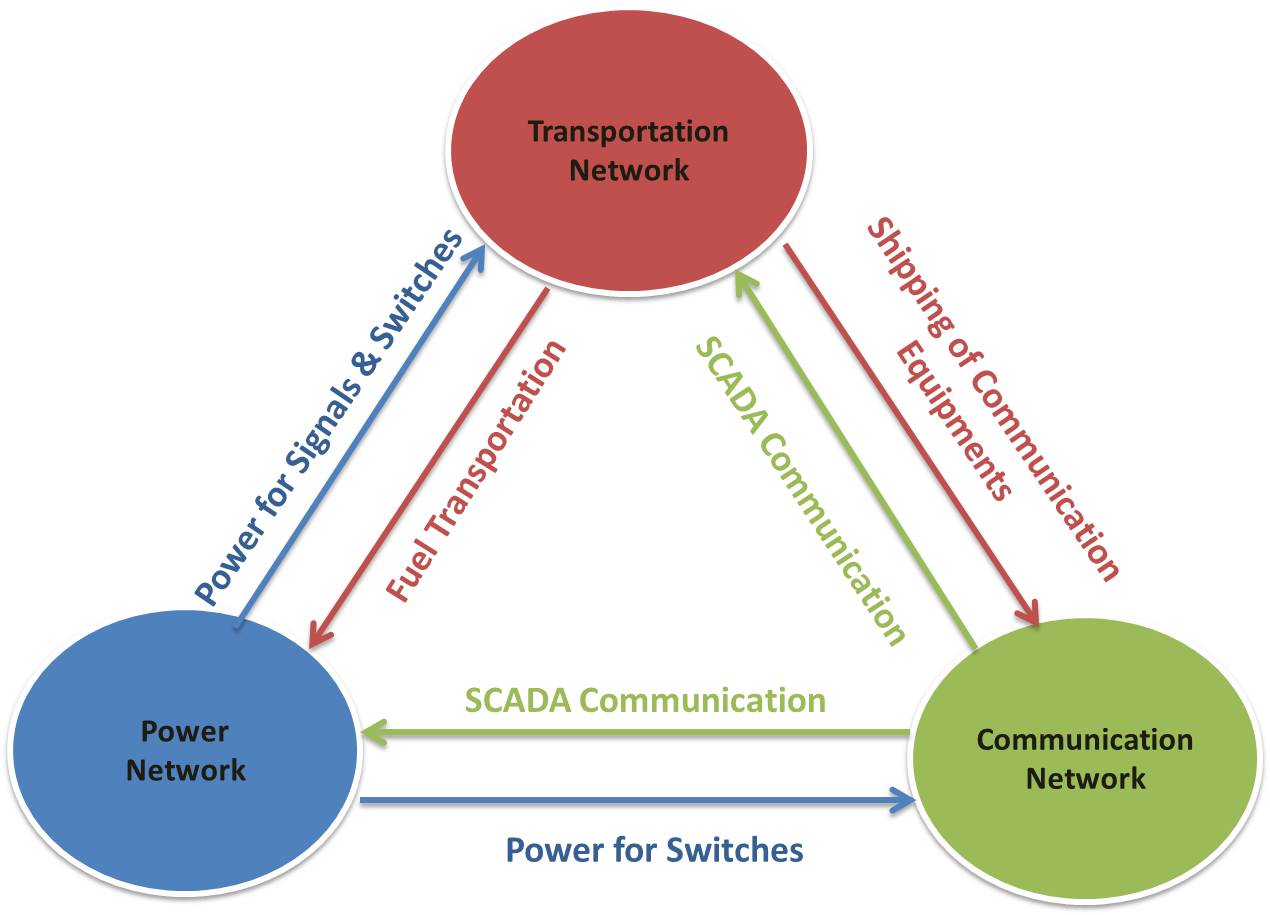}
    \caption{Interdependency between power, communication and transportation infrastructures}
    \label{fig:fig5}
\end{figure}

In the last few years there has been an increasing awareness in the research community that the critical infrastructures of the nation do not operate in isolation.  In fact, they are closely coupled with other infrastructures such that the well being of one infrastructure depends heavily on the well being of another. As an example, consider the interdependent relationship between the power, communication, and transport networks as shown in Figure \ref{fig:fig5} \cite{rinaldi2001}. If we focus exclusively on the power and communication networks we observe that entities of the power grid, such as the Supervisory Control and Data Acquisition (SCADA) systems, that control power stations and sub-stations, are dependent on the communication network to receive their operational commands. While entities of the communication network, such as routers and cell towers, are dependent on the power grid to remain operational. Compounding the complexity of analysis of this symbiotic relationship between the two networks, is the effect of cascading failures across these networks. For instance, not only can entities of the power networks, such as generators and transmission lines, trigger a power failure, but also  communication network entities, such as routers and optical fiber lines, can trigger failures in the power grid. Thus, it is essential that the interdependency between different types of networks be understood well, so that preventive measures can be taken to avoid cascading catastrophic failures in such multi-layered network environments.


With the continued focus for developing realistic failure propagation models that aid in analyzing, and mitigating the effects of cascading faults across the entities of the multi-layered network, several failure propagation models have been studied that address the interdependency relationship between power, and communication networks \cite{buldyrev2010,rosato2008,thai2013,modiano2013}, and space based networks \cite{castet2013}.

In this chapter we present a survey of the existing interdependency models for critical infrastructure networks that have been proposed in the recent literature. In Section 2, we present the models and draw attention to some of their limitations. In Section 3 we outline the considerations that need to be taken into account for capturing the complex interdependency that exists between power grid and communication networks in the real world. In Section 4 we propose an alternative model that overcomes some of the limitations of existing models by capturing the interdependency between the networks using a combination of conjunctive and disjunctive relations. Finally, in Section 5 we present concluding remarks.

\section{Interdependency Models}

\subsection{Buldyrev et al. Interdependency Model}

Motivated by the electricity blackout in Italy (2003) Buldyrev et al. \cite{buldyrev2010} proposed a cascading failure model for interdependent networks.

\begin{figure}[h]
\label{triangle}
  \centering
    \includegraphics[width=0.75\linewidth]{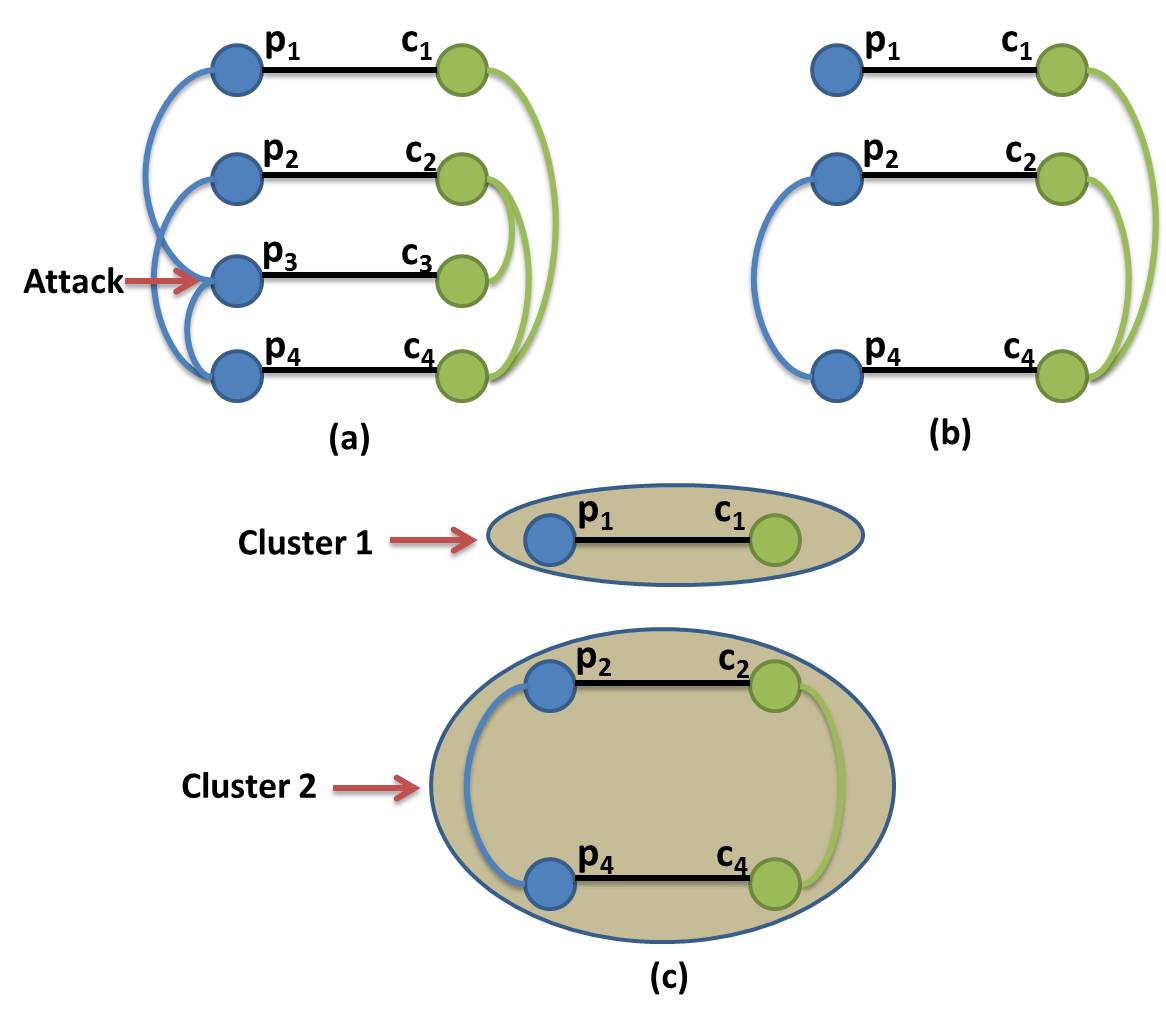}
		\caption{The interdependent network shown consists of power network nodes $p_{1}$, $p_{2}$, $p_{3}$ and $p_{4}$ and communication network nodes $c_{1}$, $c_{2}$, $c_{3}$ and $c_{4}$. Blue and green edges represent intra links in power and communication network respectively and black edges represent the interdependency (inter links). In the respective figures the cascading failure is demonstrated as follows --- (a) Node $p_{3}$ is attacked, (b) the node $p_{3}$ and its intra links are removed along with its interdependent node $c_{3}$ and its intra link, (c) the intra link $(c_{1},c_{4})$ is removed using the cascade process defined. And finally, the steady state is reached comprising of two mutually connected clusters with cluster 2 as the largest mutually connected cluster.}
    \label{fig:fig1}
\end{figure}

The power and communication infrastructures can be represented as networks. These networks are depicted as two connected graphs $P$ (for power network) and $C$ (for communication network) with same number of nodes. To represent the interdependency between the networks, bidirectional links between $P$ and $C$, termed as $P \leftrightarrow C$ edges, are considered with every node in each graph connected to exactly one node in the other graph as shown in Figure \ref{fig:fig1}(a). These bidirectional links represent the interdependency relationship that a node in the power network is dependent on exactly one node in the communication network and vice-versa. Thus capturing the fact that a failure of a node in the power (communication) network causes the corresponding node in the  communication (power) network to fail. Hence the interdependent power and communication infrastructure can be represented as a network consisting of graphs $P$ and $C$ and $P \leftrightarrow C$ edges.\



Failures are considered in the model when a fraction of the nodes from any of the two graphs $P$, or $C$ are removed. Upon the introduction of a failure in the graph $P$, the failed nodes are removed and correspondingly, the nodes in the graph $C$ that are connected via $P \leftrightarrow C$ edges to the attacked nodes are also removed. Parallel to the node removals, any edge within graph $P$ or $C$, or $ P \leftrightarrow C$ edges that do not have one node at each end point are also simultaneously removed.


The cascade now proceeds as follows. In the first stage, the set of connected components in the graph $P$ is defined as $p_{1}$ clusters. The set of $C$ nodes connected to the $p_{1}$ clusters by $P \leftrightarrow C$ edges are termed as $c_{1}$ sets. Any edges in graph $C$, that connects these $c_{1}$ sets are removed. The set of connected components in graph $C$ after this removal of edges are defined as $c_{2}$ clusters. In the second stage using same procedure as that to find the $c_{2}$ cluster and $c_{1}$ sets, $p_{2}$ sets (from $c_{2}$ clusters and $p_{3}$ clusters are obtained. In subsequent stages this cascade process then oscillates between the two graphs until a {\em steady state} is reached when no further removal of edges in the graphs are possible. At the steady state, the interdependent network consists of {\em mutually connected clusters}. Each mutually connected cluster consists of nodes having the properties (a)  the nodes in graphs $P$ and $C$ are completely connected, (b) each of these nodes which belong to the graph $P$ ($C$) has $P \leftrightarrow C$ edge with graph $C$ ($P$). Note that there exists no intra-links between any of the mutually connected clusters. An example demonstrating this cascading process is shown in Figure \ref{fig:fig1}.

The {\em largest mutually connected cluster} is defined as the cluster having the maximum number of nodes. Given a fraction $1-p$, ($0 \le p \le 1$) of nodes that are removed from the interdependent network (due to a failure), the ratio $P_{\infty}$ defines the number of nodes in the largest mutually connected cluster at the steady state, as compared to the initial number of nodes in the network. For the purpose of simulation and study, the power and communication networks are considered as, coupled scale free, Erdos Reyni \cite{erdos1959}, and random networks. Different values of $P_{\infty}$ were computed by varying the values of $p$, and the size of the network. It was observed that, above a percolation threshold $p_{c}$, the value of $P_{\infty}$ changes from the neighborhood of zero to the neighborhood of one for a given network size. From this observation the authors infer that when the fraction of failed node is below $1-p_{c}$ of the original number of nodes, the largest connected cluster has a size approximately equal to the size of initial pre-failure network. The percolation threshold $p_{c}$ for Erdos Reyni networks is validated by analytical results.


In subsequent papers, Buldyrev et al. extend their work from their original cascading failure model (as discussed above), to interdependent networks with directional dependency \cite{buldyrev2011a}, and interdependency between more than one network \cite{buldyrev2011b}.



One noticeable shortcoming of this model proposed by Buldyrev et al. is that it does not distinguish between nodes in either network as separate entities. Nodes in the power network may be functionally separate entities such as power plants, sub-stations, and load nodes. Similarly, nodes in the communication network may be functionally separate entities such as cell towers, and routers. When separate entities of the network are considered, the proposed cascading model may not work in the same way as assumed by the authors, and also the dependency relationship of one type of entity to the other may not be able to be captured with this model. Another potential drawback to this model is for the functionality of the mutually connected cluster. The mutually connected clusters generated after the cascade may not be completely functional because of the physical limitations of the network \cite{gil2012}. For example, the nodes from the power grid in a mutually connected cluster may not be able to provide sufficient power to the nodes in the communication network due to the limits on the power generation capacities. Thus, it would be wrong to assume that the residual mutually connected clusters continue to be functional after a cascade simply because they remain connected.


\subsection{Rosato et al. Coupling Model}

In \cite{rosato2008}, Rostato et al. model the power flow in the power grid, and the data flow in the communication network separately. They then analyze the effect of failures in the communication network, caused by failures in the power grid using a coupling model between the two infrastructures. Their analysis of the failure propagation is performed on the backdrop of the Italian high voltage electric transmission network (HVIET), and the high-bandwidth backbone of the Italian Internet network (GARR). Data for both the networks were gathered from documentation available in the public domain.


For modeling the power network, the HVIET network is represented by an undirected graph consisting of three type of vertices, namely, source nodes (nodes that supply power to the network), load nodes (nodes that draw power out of the network), and junction nodes (which neither draw nor supply power to network, but act as relays). The edges of the graph corresponds to the transmission lines. The power flow dynamics in the power grid relies on the DC power flow model as given by \cite{wood2012}. At every occurrence of a failure of one or more nodes, or transmission lines (edges), the power flow dynamics are recalculated using this model. It is to be noted here that the DC power flow model considers the physical constraints pertaining to the maximum power flow possible over a transmission line while computing the minimum load re-dispatch (reducing the power drawn out by the load nodes) after a failure. The authors define the quality of service (QoS) of the power network as the ratio of the change in the total power drawn by the load nodes after the failure event, as compared to the total power drawn by the load nodes before the failure event.


For modeling the communication network, the GARR network is represented as a graph consisting of high-bandwidth backbone links as edges, and the Italian universities and research institutions as nodes. For computing the total amount of traffic inflow into the network, the probability that a node generates a packet $\lambda$, ($ 0 \le \lambda \le 1$) is considered at each time step. For each generated packet a random node is chosen as its destination. A probabilistic packet routing model is considered along the lines of \cite{echenique2004} for sending the packets to their intended destinations. The average delivery time is defined as the average of the packet transmission time from source to destination over all packets delivered correctly within a particular time interval. The average delivery time is then used as a metric to define the efficiency of the network for a given value of $\lambda$.


The coupling between the two networks is achieved by associating a node from the communication network to the closest load node from the power network (Euclidean distance). Note that this coupling is one directional, that is, for a node to be operational in the communication network it is dependent on a node from the power network, but the power network node is not dependent on the communication node for its survival. In a failure event, if a load node $i$ that was initially extracting power $P_{i}^{0}$ units, now extracts $P_{i}$ units of power after the subsequent load re-dispatching process. The communication nodes coupled to $i$ remain operational as long as the value of $P_{i}$ is greater than or equal to $\alpha P_{i}^{0}$, ($ 0 \le \alpha \le 1$). The coefficient $\alpha$ is termed as the strength of coupling between the two networks.


The authors then use the above coupling model to analyze and simulate the effect of random link failures in the power network for a fixed parameter of $\alpha$ (taken as $\alpha=0.75$). The main insight of their simulation is that even with small failure events in the power (HVIET) network (small with respect to number of transmission lines failed), the communication (GARR) network can get completely disconnected.


The individualized modeling of the power and communication network done by Rostato et al. is realistic to a point, but the coupling model reflects only a one way dependency model and fails to represent the interdependency that exists between power and communication networks of today. This shortcoming may prohibit the accurate cascading failure scenarios when the faults originate from the communication network and cascade through to the power network.


\subsection{Nguyen et al. Interdependency Model}

\begin{figure}[h]
\label{triangle}
  \centering
    \includegraphics[width=0.75\linewidth]{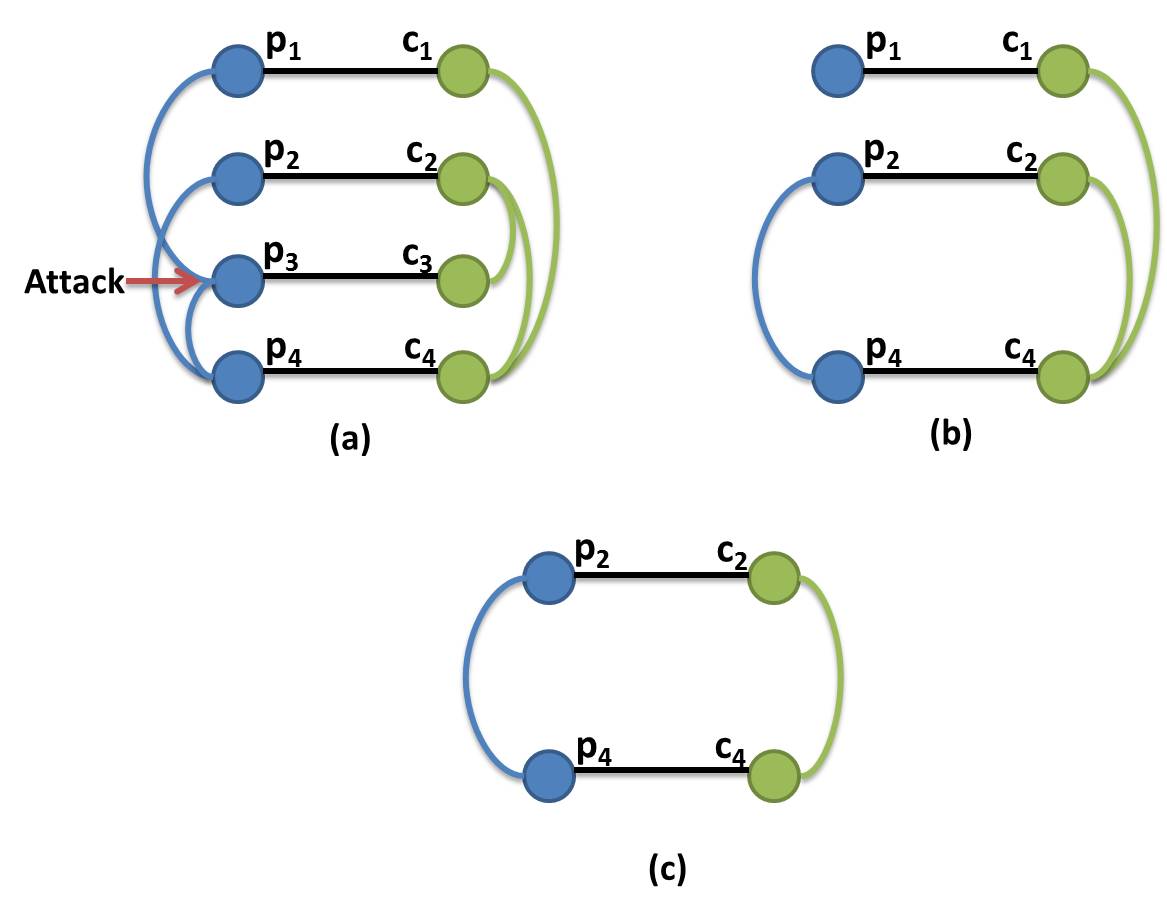}
    \caption{Power Network consisting of nodes $p_{1}$,$p_{2}$,$p_{3}$,$p_{4}$ and Communication Network consisting of nodes $c_{1},c_{2},c_{3},c_{4}$. Blue and green edges represents intra links in power and communication network respectively and black edges represent the interdependency (interlinks). (a) The node $p_{3}$ is attacked. (b) The intra links of node $p_{3}$ are removed due to its failure along with its interdependent node $c_{3}$ in communication network and all its associated intra links. (c) The node $p_{1}$ fails as it is disconnected from the the largest connected component in the power network. The steady state is reached with nodes $p_{2}$ and $p_{4}$ in power network as functional nodes after the failure event.}
    \label{fig:fig2}
\end{figure}

In \cite{thai2013}, Nguyen et al. propose a cascading model in similar lines of \cite{buldyrev2010}, and address the problem for identifying the critical nodes in an interdependent network. In their model, the power network, and communication network are considered as graphs $G_{s}=(V_{s},E_{s})$ and $G_{c}=(V_{c},E_{c})$, and the interdependency is represented by an unidirectional edge set $E_{sc}$ that connect vertices from set $V_{s}$ with set $V_{c}$ in a composite graph containing this edge set, and both the power, and communication networks graphs. A failure due to a dependency relation is outlined by the assumption that, not only do the failed node(s) cease to operate, but also the nodes connected to the failed nodes via edges from the edge set $E_{sc}$ also become non-operational. Failures propagate in the following way: the failed nodes and the incident edges to these nodes that belong to $G_{s}$ (power network), and $G_{c}$ (communication network) are removed to generate $G'_{s}$ and $G'_{c}$ respectively. Then, the largest connected components $L_s$ and $L_c$ are computed for the graphs $G'_{s}$ and $G'_{c}$. Any node $n_s \in G'_{s}$ that does not belong to $L_s$, and any node $n_c \in G'_{c}$ that does not belong to $L_c$ are considered non-operational. Failures due to the dependency relations are simultaneously considered, and propagation ensues until a {\em steady state} is reached when no further nodes in either network can fail. An example showing this failure propagation is shown in Figure \ref{fig:fig2}.


Using the above defined failure propagation model, the authors consider the problem of identifying a set of critical nodes in the power network of size less than a positive integer $k$, such that at the steady state the size of the largest connected component in the power network is minimized. The authors show that this problem is NP-complete by reduction from the decision version of the Maximum Independent Set problem, and infer that this problem is in-approximable within a bound of $2-\epsilon$. Three greedy approximation algorithms are proposed by the authors for approximating the solution to this problem in polynomial time, namely, {\em Maximum Cascade} (Max-Cas),  {\em Iterative Interdependent Centrality} (IIC), and  {\em Hybrid}.


The authors perform an extensive simulation of the proposed algorithms using three different power network, and communication network data sets. The data sets considered were (i) US Western States power network, and a synthetic scale free communication network with an exponential factor, $\beta = 2.2$, (ii) Synthetic scale free power network with $\beta = 3.0$, and a synthetic scale free communication network with $\beta = 2.2$, and (ii) Scale free power and communication networks with the same $\beta = 2.6$. For each of the simulations the interdependency relationship between the two networks were setup using a random weighted permutation of nodes of the two networks.


In the simulations it was observed that the Hybrid algorithm takes lesser time and has better performance bounds than the other two algorithms. In the process of the simulations, it was observed that when interdependent systems are loosely connected they are more vulnerable to failure. Their observations also included that sparse interdependent networks are more vulnerable to cascading failures. This was observed from simulations carried out by varying the exponential factor of the scale free communication network, while keeping the exponential factor of the power network, and the total number of nodes constant. The simulations carried out by the authors by varying the total number of nodes of both the networks, while keeping a fixed exponential factor of the considered scale free networks, showed that large networks are more vulnerable to cascading failures.


The observable shortcomings of this model are similar to the drawbacks discussed above for the model proposed by Buldyrev et al. \cite{buldyrev2010}. Without the distinction of nodes in the networks into separate entities, such as power plants, and substations, for the power network, and cell towers, and routers for the communication network, the failure cascading model may not represent the workings of real world networks. Thus hindering the analysis, and mitigation of faults caused by cascading failures in multi-layer critical infrastructure networks.


\subsection{Parandehgheibi et al. Interdependency Model}

\begin{figure}[h]
\label{triangle}
  \centering
    \includegraphics[width=0.5\linewidth]{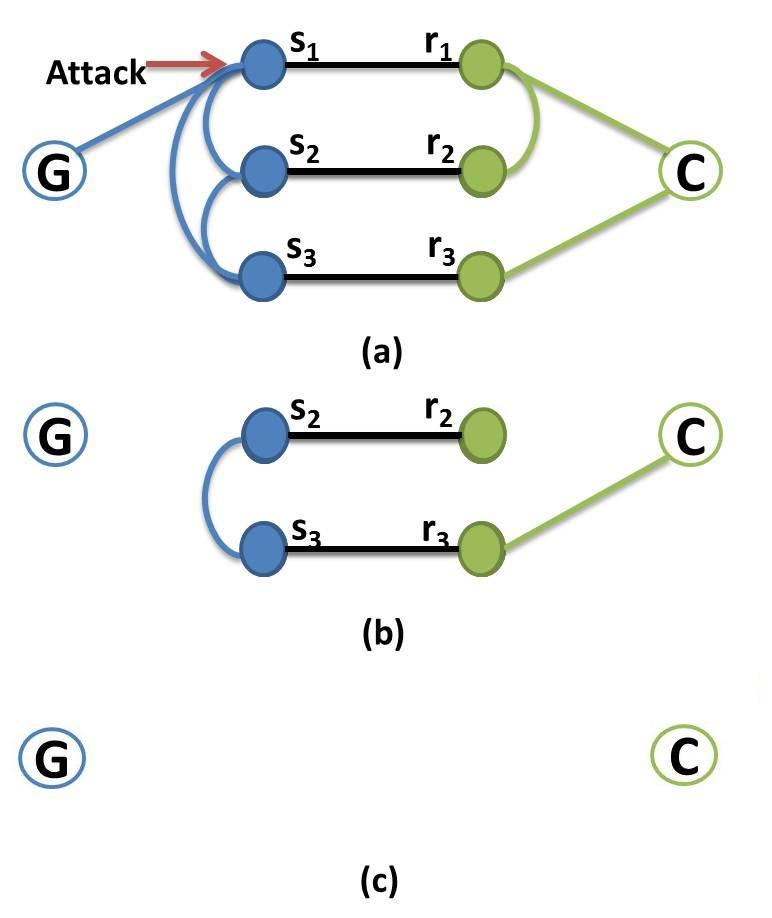}
    \caption{The power network consists of a generator $G$ and substations $s_{1},s_{2},s_{3}$ and communication network consists of control center $C$ and routers $r_{1},r_{2},r_{3}$. Blue edges denotes the power network edges (composed of transmission lines) and green edges denotes the communication network edges (composed of communication links). Black edges denotes the interdependency between substation of power network and routers in communication network. (a) The substation $s_{1}$ is attacked. (b) Failure of substation $s_{1}$ results in removal of all power network edges incident on $s_{1}$ and failure of interdependent router $r_{1}$ and removal of communication network edges incident on it. (c) Substations $s_{2}$, $s_{3}$ and routers $r_{2}$, $r_{3}$ fails and hence are removed as they do not satisfy both the properties for being being functional as mentioned. The edges incident on these substations and routers are subsequently removed. The resultant interdependent network after the failure consists of two autonomous nodes $G$ and $C$.}
    \label{fig:fig3}
\end{figure}

Parandehgheibi et al. \cite{modiano2013} also consider the power and communication infrastructure networks to analyze the effect of cascading failures on these interdependent networks. In their model, the power network graph $P=(V_p, E_p)$ consist of vertices $V_p$ representing the generators, and substations, and edges $E_p$ representing the transmission lines. Similarly, the communication network graph $C=(V_c, E_c)$ consist of vertices $V_c$ representing the control centers, and routers, edges $E_c$ representing the communication lines. In the graphs, it is assumed that nodes represented by generators, and control centers, are {\em autonomous}, i.e. these nodes operate independently without any dependency on any other node across both the networks. In this model, dependency between network entities is represented by coupling the routers, and substations with edges $E$ (directed or undirected), in a composite graph of $G=(V, E, E_p, E_c), V=V_p \cup V_c$. Whether a node of this composite graph $G$ is operational or not is defined by the following functional rules: If the node represents a substation, it remains operational as long as, (i) there exists a path between the substation and a generator via the power network edges $E_p$, and (ii) there exists a path between the substation and a router (to receive control signals) via edges of $E$. If the node represents a router, it remains operational as long as, (i) there exists a path between the router and a control center via the communication network edges $E_c$, and (ii) there exists a path between the router and a substation (to receive power) via edges of $E$. Lastly, if the node represents a generator, or a control center, it remains continuously functional. At the time of the initial failure (due to a possible attack, or fault), the failed nodes, or edges are removed from the graph $G$. The failure propagation is then represented in the model by iteratively removing the failed nodes and all their incident edges from graph $G$ that do not satisfy the aforementioned functional rules. This propagation continues until a {\em steady state} is reached when no further removals of nodes, or edges are necessary. An example of the described failure propagation is illustrated in Figure \ref{fig:fig3}.

Keeping this failure model as their basis, the authors consider the problem of selection of the minimum number of non-autonomous nodes (substations, and routers), that need to be removed from the graph $G$, such that the resulting graph  generated at the steady state contains no non-autonomous nodes. The authors term this problem as the {\em Node-MTFR} (minimum total failure removal) problem. They also identify another similar problem {\em Edge-MTFR}, that concentrates on the selection of the minimum number of edges of $G$ such that the resulting graph generated at the steady state contains no non-autonomous node.

For solving these two problems the authors assume that the power network graph $P$, and the communication network graph $C$, are each star topology graphs. For the power network, the substations are directly connected to a generator without any connections between any other substations, i.e for all edges $(u,v) \in E_p$, node $u$ represents a substation, and node $v$ represents a generator. Similarly for the communication network the router are directly connected to a control center without any connections between any other routers, i.e for all edges $(u,v) \in E_c$, node $u$ represents a router node, and node $v$ represents a control center node. The authors now proceed to analyze the problem from the perspective of a bipartite graph, where the nodes in the bipartite graph comprise of the substations of the power network, and routers of the communication network (the nodes representing generators, and control centers are ignored). The edges of this bipartite graph is the set of dependency relations represented by edge set $E$, of graph $G$. The authors  analyze this problem from two interdependency perspectives, namely, unidirectional interdependency, and bi-directional interdependency.


For unidirectional dependency, the Node-MTFR problem is shown to be NP-complete by reduction from the Feedback Vertex Set problem, and an optimal solution is proposed by an integer linear program (ILP). A greedy approximation algorithm is also proposed for this problem and its solution is compared with the optimal solution obtained from the ILP. The authors also prove that Edge-MTFR problem for unidirectional interdependency is NP-complete by reduction from  the Feedback Edge Set problem.


For bidirectional interdependency, the authors show that the Node-MTFR problem corresponds to a minimum vertex cover problem for bipartite graphs, and using Konig's Theorem, show that this problem is equivalent to the maximum matching problem for bipartite graphs which has a known polynomial time solvable algorithm \cite{ahuja1993}. Thus showing that the Node-MTRFR problem for bidirectional interdependency is polynomially solvable. The authors also observe that for the Edge-MTFR problem with bidirectional interdependency all the edges of the bipartite graph must necessarily be removed, as any existing edge would denote the existence of operating non-autonomous nodes.


For the purpose of experimentation and simulation, the authors use the Italian communication and power network data obtained from \cite{rosato2008}. To preserve the star topology configuration for the power and communication networks,  only substations directly connected to the generators, and routers directly connected to control centers are considered. Unidirectional dependency between the substations and routers is established by assuming that a substation receives  control signals from the nearest router, and a router receives power from the nearest substation. Using this setup the simulation is carried out to find the minimum number of nodes representing routers and substations that need to be removed, such that all non-autonomous nodes are removed from the graph (Node-MTFR). The experimental results showed that the north-western part of Italy is acutely vulnerable as removal of just three routers results in the failure of all substations and remaining routers.



A possible drawback to this model is that this model is able to represent dependencies that are in disjunctive form, for example, a sub-station survives as long it has a connection to a router. However, if there is a need to model a conjunctive dependency among network entities this model may not be adequate, for example, a scenario where a sub-station survives only when it is connected to two routers. In the real world, it is highly likely that entities in either the power or communication network have such conjunctive dependency amongst other entities, which this model may not be able to adequately represent. Another possible shortcoming of this model is the number of types of power, and communication entities that this model considers. For instance, in a real world communication network there may be communication entities such as cell towers whose survivability may have to be modeled very differently than the way routers are modeled. In the proposed model if support for additional entities are included that have different functional rules, it is not clear how this model will be able to accommodate them.



\subsection{Castet et al. Interdependency Model}
In \cite{castet2013}, Castet et al. develop a model for survivability analysis of networks with heterogeneous nodes (nodes that can perform more than one function), and apply their approach to space-based networks. The authors propose that heterogeneous networks can be modeled as interdependent multi-layer networks, thus enabling survivability analysis of these networks. They assert that in this approach, the multi-layer aspect captures the common functionalities across the different nodes (by construction of homogeneous sub-networks), and the interdependency aspect captures the physical characteristics of each node in the network.

In this paper the authors focuses on space-based networks (SBNs). In SBNs, each network entity (space-craft), may perform more than one function. SBN's operate by physically distributing functions in multiple orbiting space-crafts that are wirelessly connected to each other. The SBNs architecture allows the sharing of resources on-orbit, such as data processing, data storage, and downlinks among the network entities. In this study, Castet et al. attempt to assess their proposed approach of modeling heterogeneous networks as interdependent multi-layer networks on SBNs, and benchmark the survivability of a fractionated SBN architecture, against that of a traditional monolith spacecraft.

To represent the heterogeneous SBN as a multi-layer interdependent network the authors define the following terms:
\begin{itemize}
	\item {\em Super-Node}: A network entity that supports multiple functionalities
	\item {\em Node}: Component of a super-node that represents a single functionality of that super-node
	\item {\em Layer}: Set of nodes with the same functionality
	\item {\em Intra-Layer Link}: A link between two nodes in the same layer. The link can be directed (when one node is providing a resource and the other is receiving), or undirected (both provide, and receive resources)
	\item {\em Networked Layer}: A network possessing intra-layer links
	\item {\em Inter-Layer Link}: A directed link that captures the inter-dependency between functionalities (nodes) within a super-node. Specifically, this link implies the (directed) propagation of failure from one node to the other.
\end{itemize}

In their model two types of inter-layer links are considered that represent the two types of failure propagation possible in the model: (i) Inter-links for the {\em kill effect} failure propagation, defined by the propagation rule as follows: When a node fails, all nodes that have an incoming inter-link of this type from the failed node immediately fail, and (ii) Inter-links for the {\em precursor effect} failure propagation, defined by a conditional propagation rule as follows: When a node fails, and all the nodes with incoming intra-links to this failed node have also failed, all entities that have an incoming inter-link of this type from the failed node fails. This type of inter-link implicitly implies that as long as a super-node has access to a particular functionality, either from its own resources or from another super-node, all nodes in the super-node dependent on this functionality survive.

\begin{figure}[h]
\label{triangle}
  \centering
    \includegraphics[width=0.55\linewidth]{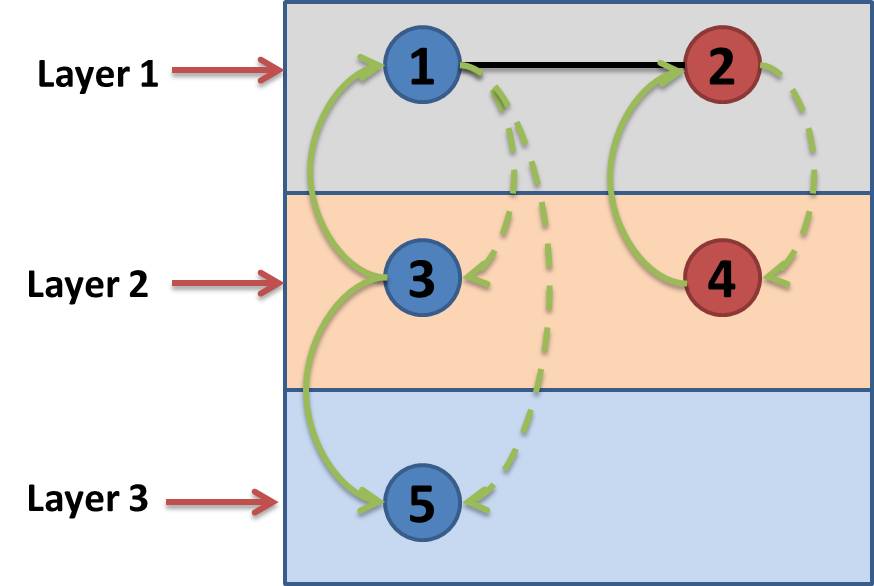}
    \caption{Interdependent space based network consisting of three layers represented by graphs $G_{1}=(\{1,2\},\{(1,2),(2,1)\})$, $G_{2}=(\{3,4\},\emptyset )$, $G_{2}=(\{5\},\emptyset )$. Edge set $E_{k}=\{(3,1),(3,5),(4,2)\}$ and edge set $E_{p}=\{(1,3),(1,5),(2,4)\}$. If node $3$ fails, nodes $1$ and $5$ immediately fail ({\em kill effect}). If node $1$ fails then nodes $3$ and $5$ don't fail unless node $2$ also fails ({\em precursor effect}).}
    \label{fig:figure4}
\end{figure}

Figure \ref{fig:figure4} demonstrates the propagation rules and represents a sample SBN as an interdependent multi-layer network $N$ defined by $N(G_{1},..., G_{L},E_{k},E_{p})$, where:
\[
\begin{cases}
L & is\: the\: number\: of\: layers\: each\: numbered\: sequentially\: from\:1\: to\: L\\
 & G_{1},....,G_{L}\: are\: the\: graphs\: on\: each\: layer:\\
 & \quad\forall l\in[1,...,L],G_{l}=(V_{l},E_{l})\: with:\\
 & \qquad\qquad\begin{cases}
V_{l} & is\: the\: set\: of\: n_{l}\: nodes\: in\: G_{l}\\
E_{l} & is\: the\: set\: of\: intra-layer\: links\: in\: G_{l}
\end{cases}\\
E_{k} & is\: the\: set\: of\: inter-layer\: links\: representing\: the\:"kill\: effect"\\
E_{p} & is\: the\: set\: of\: inter-layer\: links\: representing\: the\:"precursor\: effect"
\end{cases}
\]

To analyze the survivability of an interdependent multi-layer network using the above network representation, and propagation rules, the authors carry out the following steps: (i) Generate the time to failure for each node and intra-layer link, (ii) propagate failures through inter-layer links for the kill effect, (iii) propagate failures through inter-layer links for the precursor effect, and (iv) combine all failure propagation effects to obtain the probability of failure of each node. Random times to failure for the nodes were generated using cumulative distribution functions representing the failure behavior of each node. Since links between two space-crafts (super-nodes) is established through a wireless unit, a two step process was followed for generating the times to failure for the intra-layer links: (i) times to failure of the wireless units on each spacecraft was generated using predetermined cumulative distribution functions, (ii) times to failures for each intra-layer link was generated by taking the minimum of the time to failures of the two associated wireless units.

For simulation and study, the authors apply their model into three different SBN scenarios. In their first scenario they consider three different space network architectures. The first architecture considered consists of a traditional monolith spacecraft with three subsystems (or layers), namely, \textit{Telemetry Tracking and Command} (TTC), \textit{supporting subsystems}, and \textit{payload}. The second architecture consists of two space based networks, one of them a traditional monolith spacecraft, while the other spacecraft consists of two subsystems --- \textit{TTC} and \textit{supporting subsystems}. The two spacecrafts shares their TTC subsystems, i.e. a TTC redundancy is introduced, through a wireless link. This architecture is shown in Figure \ref{fig:figure4} with layer $1$,$2$ and $3$ denoting subsystems TTC, supporting subsystems, and payload respectively. A third architecture is considered which is comprised of the monolith spacecraft, and two spacecrafts having two subsystems --- \textit{TTC} and \textit{supporting subsystems}. These three spacecrafts share there TTC subsystems, i.e. there is a higher degree of TTC redundancy, through wireless links. Wireless links in the second and third spacecraft architecture are assumed to be perfect. The distribution of probability of unavailability (failure) of TTC subsystem with time, identified as a major spacecraft unreliability factor in \cite{castet3}, is obtained from \cite{castet2}. The probability of unavailability of the payload subsystem over time, for the three spacecrafts is computed considering the failure of the TTC subsystem using a Monte Carlo Simulation. The simulation results showed that for a given time, increasing the redundancy of the TTC subsystems reduces the probability of unavailability of the payload. However, it was observed that the percentage of this reduction is not linear with the redundancy introduced.


The second scenario was aimed to study the impact of wireless link failure. A Weibull distribution is considered for probability of unavailability of wireless link failure with time. The parameters of Weibull distribution are set such that the wireless link has a probability of $0.5$ to fail after 15 years. Simulations were carried out to compute the probability of unavailability of payload for the second architecture of the first application with the given wireless link failure distribution. The result is compared with the first and second architecture with perfect wireless link (the previous scenario). Compared with the first scenario, it was observed that for the second architecture the probability of unavailability of payload increases with time when wireless link failure is considered. At a given point in time, it surpasses the probability of unavailability of monolith spacecraft thus negating the effect of a TTC redundancy. The conclusion that can be drawn from these observations are that failure behavior of wireless links is a critical consideration to analyze the advantage of space based networks with TTC redundancy, over adoption of traditional monolith space crafts.

In the third scenario the authors consider a more complex space based network by including two new subsystems into the traditional monolith spacecraft. The new subsystems included are a Control Processor (CP) subsystem (the main computer of the spacecraft), and a Data Handling (DH) subsystem (handling exchange and storage of data). Another space craft is considered with all the subsystems as stated except the payload. These two spacecrafts share DH, TTC and CP subsystems, thus introducing redundancy. The resources are shared via wireless links. Hence the space based network represented by this architecture has 5 layers with 3 networked layer. The distribution of probability of unavailability of these subsystems with time is obtained from \cite{castet2}. Assuming perfect wireless link, a Monte Carlo simulation is carried out to compute the probability of unavailability of payload with time. The simulation result is compared with traditional monolith spacecraft, and the second spacecraft architecture's (from the first scenario) payload failure distribution. It is observed that after 15 years this architecture reduces the risk of failure by $20.5 \%$ over the monolith spacecraft. This makes way to draw a conclusion that this architecture has greater improvement in reduction of failure over monolith spacecraft, than by only introducing TTC redundancy (as considered in first scenario).

\section{Limitations of Current Modeling Approaches and Possible Solutions}

As discussed in the previous section, significant efforts have been made in the research community in the last few years to develop an appropriate model of interdependency between the entities of a multi-layer critical infrastructure network \cite{buldyrev2010, buldyrev2011a, buldyrev2011b, rosato2008, peeta1, castet2013, modiano2013, chen2012, thai2013,Sen091,Zus11}. Unfortunately, many of the proposed models are overly simplistic in nature and as such they fail to capture the complex interdependency that exists between power grid and communication networks. As noted in section 2.1, the highly cited paper due to Buldyrev et al. \cite{buldyrev2010}, assume that every node in one network can depend on one and only one node of the other network. Obviously, this assumption is not valid in an interdependent power-communication network that spans countries and continents,  Even the authors in a follow up paper \cite{buldyrev2011b} recognize that the assumption may not be valid in the real world and a single node in one network may depend on more than one node in the other network and vice-versa. A node in one network may be functional (``alive'') as long as one supporting node on the other network is functional.

Although this generalization can account for {\em disjunctive dependency} of a node in the $A$ network (say $a_i$) on more than one node in the $B$ network (say, $b_j$ and $b_k$), implying that $a_i$ may be ``alive'' as long as either $b_i$ or $b_j$ is alive, it cannot account for {\em conjunctive dependency} of the form when {\em both $b_j$ and $b_k$} has to be alive in order for $a_i$ to be alive. In a real network the dependency is likely to be even more complex involving both disjunctive and conjunctive components. For example, $a_i$ may be alive if (i) $b_j$ {\em and} $b_k$ {\em and} $b_l$ are alive, {\em or} (ii) $b_m$ {\em and} $b_n$ are alive, {\em or} (iii) $b_p$ is alive. The graph based interdependency models proposed in the literature \cite{buldyrev2011a, rosato2008, peeta1, castet2013, modiano2013, thai2013} including \cite{buldyrev2010, buldyrev2011b} cannot capture such complex interdependency between entities of multi-layer networks. In order to capture such complex interdependency, we propose recently a new model of interdependency using Boolean logic \cite{Sen14}. In the following, we briefly describe this model.

We outline the model for an interdependent network with two layers. However, the concept can easily be generalized to deal with networks with more layers. Suppose that the network entities in layer 1 are referred to as the $A$ type entities, $A = \{a_1, \ldots, a_n\}$ and entities in layer 2 are referred to as the $B$ type entities, $B = \{b_1, \ldots, b_m\}$. If the layer 1 entity $a_i$ is {\em operational} if (i) the layer 2 entities $b_j, b_k, b_l$ are operational, or (ii) $b_m, b_n$ are operational, or (iii) $b_p$ is operational, we express it in terms of {\em Live Equations} of the form $a_i \leftarrow b_jb_kb_l + b_mb_n + b_p$.  The Live Equation for a $B$ type entity $b_r$ can be expressed in a similar fashion in terms of $A$ type entities. If $b_r$ is {\em operational} if (i) the layer 1 entities $a_s, a_t, a_u, a_v$ are operational, or (ii) $a_w, a_z$ are operational, we express it in terms of {\em Live Equations} of the form $b_r \leftarrow a_sa_ta_ua_v + a_wa_z$. It may be noted that the {\em live equations} only provide a {\em necessary condition} for entities such as $a_i$ or $b_r$ to be {\em operational}. In other words, $a_i$ or $b_r$ may {\em fail} independently and may be {\em not operational} even when the conditions given by the corresponding {\em live equations} are {\em satisfied}.
A live equation in general will have the following form: \[ x_i \leftarrow \sum_{j = 1}^{T_i}\prod_{k = 1}^{t_j} y_{j, k}\] where $x_i$ and $y_{j, k}$ are elements of the set $A$ ($B$) and $B$ ($A$) respectively, $T_i$ represents the {\em number} of min-terms in the live equation and $t_j$ refers to the {\em size} of the $j$-th min-term (the size of a min-term is equal to the number of $A$ or $B$ elements in that min-term). In the example $a_i \leftarrow b_jb_kb_l + b_mb_n + b_p$, $T_i = 3$, $t_1= 3, t_2 = 2, t_3 = 1$, $x_i = a_i$, $y_{2, 1} = b_m$, $y_{2, 2} = b_p$.

\begin{table*}[h]
\begin{center}
\begin{tabular}{|c||c|}  \hline
{\bf Power Network} & {\bf Communication Network} \\ \hline
$a_1\leftarrow b_1 + b_2$ & $b_1 \leftarrow a_1 + a_2a_3$ \\ \hline
$a_2 \leftarrow  b_1b_3 + b_2$ & $b_2 \leftarrow a_1 + a_3$ \\ \hline
$a_3 \leftarrow b_1b_2b_3$ & $b_3 \leftarrow a_1a_2$ \\ \hline
$a_4 \leftarrow b_1 + b_2 + b_3$ & $--$ \\ \hline
\end{tabular}
\caption{Life Equations for a Multi-layer Network}
\protect\label{eqTbl}
\end{center}
\end{table*}
\vspace{-0.3in}
\begin{table*}[h]
\begin{center}
\begin{tabular}{|c|c|c|c|c|c|c|c|}  \hline
\multicolumn{1}{|c|}{Entities} & \multicolumn{7}{c|}{Time Steps}\\
\cline{2-8} & $t_0$ & $t_1$ & $t_2$ & $t_3$ & $t_4$ & $t_5$ & $t_6$ \\\hline \hline
$a_1$ & $1$ & $1$ & $1$ & $1$ & $1$ & $1$ & $1$ \\ \hline
$a_2$ & $0$ & $0$ & $0$ & $0$ & $1$ & $1$ & $1$ \\ \hline
$a_3$ & $0$ & $0$ & $1$ & $1$ & $1$ & $1$ & $1$ \\ \hline
$a_4$ & $0$ & $0$ & $0$ & $0$ & $1$ & $1$ & $1$ \\ \hline
$b_1$ & $0$ & $0$ & $0$ & $1$ & $1$ & $1$ & $1$ \\ \hline
$b_2$ & $0$ & $0$ & $0$ & $1$ & $1$ & $1$ & $1$ \\ \hline
$b_3$ & $0$ & $1$ & $1$ & $1$ & $1$ & $1$ & $1$ \\ \hline
\end{tabular}
\caption{Time Stepped Cascade Effect for a Multi-layer Network}
\protect\label{cascadeTbl}
\end{center}
\end{table*}
\vspace{-0.2in}
We refer to the live equations of the form $a_i \leftarrow b_jb_kb_l + b_mb_n + b_p$ as {\em First Order Dependency Relations} also, because these relations express direct dependency of the $A$ type entities on $B$ type entities and vice-versa. It may be noted however that as $A$ type entities are dependent on $B$ type entities, which in turn depends on $A$ type entities, failure of some $A$ type entities can trigger failure of other $A$ type entities, though {\em indirectly} through some $B$ type entities. Such interdependency creates a {\em cascade of failures} in multi-layered networks when only a few entities of either $A$ type or $B$ type (or a combination) fails. We illustrate this with the help of an example. The live equations for this example is shown in Table \ref{eqTbl}.

\begin{figure*}[ht]
  \subfigure[Cascading failures reach steady state after $p$ time steps]{\label{fig:cascade}\includegraphics[width=0.45\textwidth, keepaspectratio]{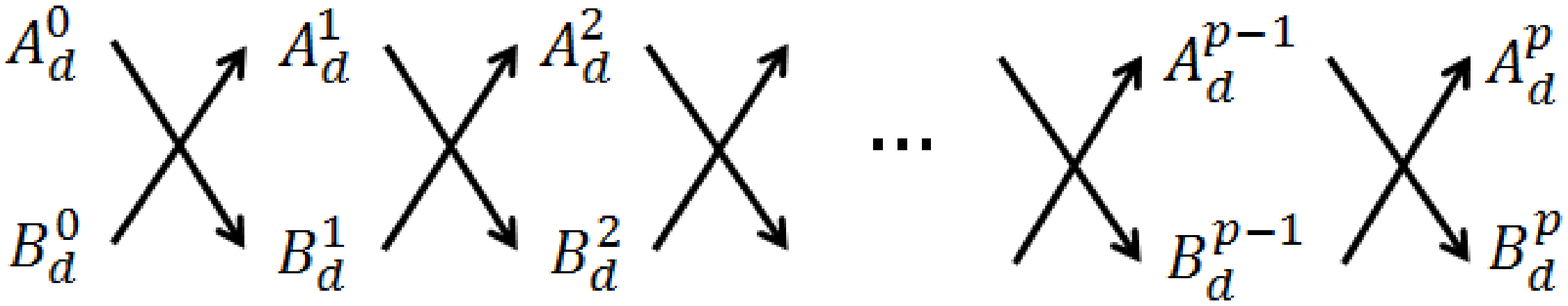}}
  \subfigure[Cascading failures as a fixed point system]{\label{fig:fixedPoint}\includegraphics[width=0.45\textwidth, keepaspectratio]{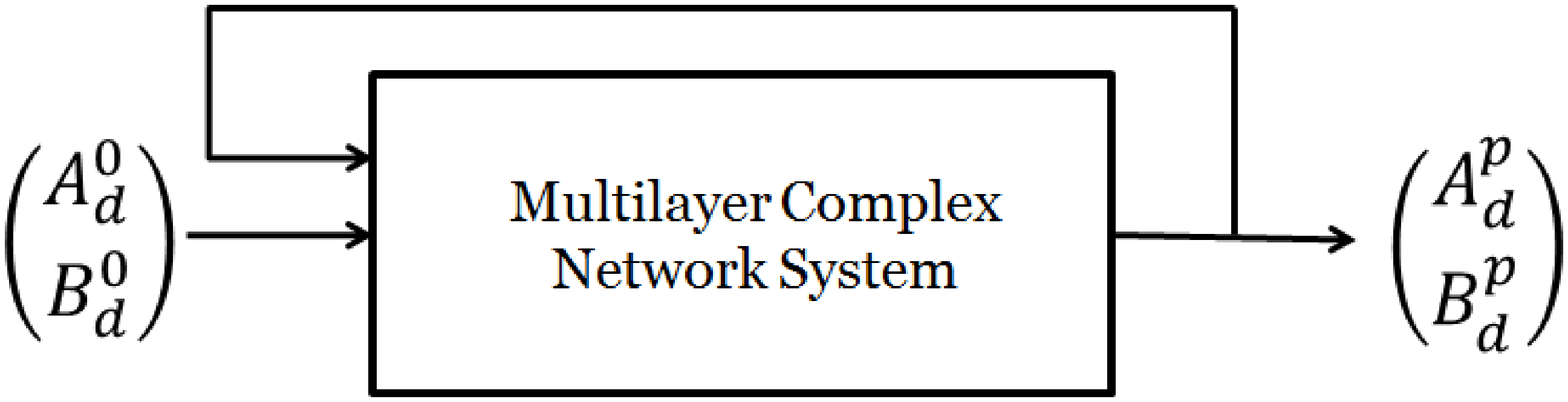}}
  \caption{Cascading Failures in Multi-layered Networks}  \label{fig:cascadeFixedPoint}
\end{figure*}

As shown in Table \ref{cascadeTbl}, in this example the failure of only one entity $a_1$ at time step $t_0$ triggered a chain of failures that resulted in the failure of all the entities of the network after by timestep $t_4$. A table entry of 1 indicates that the entity is ``dead''. In this example, the failure of $a_1$ at $t_0$ triggered the failure of $b_3$ at $t_1$, which in turn triggered the failure of $a_3$ at $t_2$. The failure of $b_3$ at $t_1$ was due to the dependency relation $b_3 \leftarrow a_1a_2$ and the failure of $a_3$ at $t_2$ was due to the dependency relation $a_3 \leftarrow b_1b_2b_3$. The cascading failure process initiated by failure (or death) of a subset of $A$ type entities at timestep $t = 0$, $A_{d}^0$ and a subset of $B$ type entities $B_{d}^{0}$ till it reaches its final steady state is shown diagrammatically in Figure \ref{fig:cascade}. Accordingly, a multi-layered network can be viewed as a ``closed loop'' control system as shown in Figure \ref{fig:fixedPoint}. Finding the steady state after an initial failure in this case is equivalent of computing the {\em fixed point} of a function $F(.)$ such that $F (A_{d}^p \cup B_{d}^p) = A_{d}^p \cup B_{d}^p$, where $p$ represents the number of steps when the system reaches the steady state.

We define a set of $k$ entities in a multi-layered network as “most vulnerable” if failure of these $k$ entities triggers the failure of the largest number of other entities. The goal of the $k$ most vulnerable nodes in multi-layered network problem is to identify this set of nodes. This is equivalent to identifying $A_{d}^0 \subseteq A$, $B_{d}^0 \subseteq B$, that {\em maximizes} $|A_{d}^p \cup B_{d}^p|$, subject to the constraint that $|A_{d}^0 \cup B_{d}^0| = k$.

The dependency relations (live equations) can be formed either after careful analysis of the multi-layer network along the lines carried out in \cite{Zus11}, or after consultation with the engineers of the local utility and internet service providers.

Utilizing this comprehensive model, we provide techniques to identify the $k$ most vulnerable nodes of an interdependent multi-layered network system in \cite{Sen14}, so that preventive measures can be taken to strengthen the network. We show that the this problem can be solved in polynomial time for some special cases, whereas for some others, the problem is NP-complete. We also show that this problem is equivalent to computation of a {\em fixed point} \cite{Fud2010} of a {\em closed loop system} and we provide a technique utilizing Integer Linear Programming to compute that fixed point. Finally, we present the efficacy of our technique using real data collected from power grid and communication networks that span the Maricopa county of Arizona in \cite{Sen14}. 

\section{Conclusion}

In order to build a robust and resilient system, a deep understanding of the complex interdependency that exists between critical infrastructures such as the power grid and the communication network is essential. Unfortunately, many of the proposed models are unable to capture such complex interdependency. In our opinion, the model proposed in \cite{Sen14} is a step in the right direction. However many problems, including the problem of model validation, still remain open. These problems will most likely draw the attention of the researchers in this domain for many years to come.


\end{document}